\documentclass[12pt,twoside,a4paper]{article}
\usepackage{qmc,epsfig}

\newcommand{\bl}{\begin{itemize}}
\newcommand{\el}{\end{itemize}}
\newcommand{\be}{\[}
\newcommand{\ee}{\]}
\newcommand{\ba}{\begin{eqnarray*}}
\newcommand{\ea}{\end{eqnarray*}}

\newcommand{\ev}[1]{E\left(#1\right)}

\newcommand{\boa}{{\bf a}}
\newcommand{\bq}{{\bf q}}

\newcommand{\ens}{{\cal E}}

\newcommand{\mod}{\mathop{\rm mod}\nolimits}

\begin{document}

\title{
{\vspace{-1em} \normalsize
\hfill \parbox{70mm}{MS-TP-01-3, IFUP-TH 2001/26}}\\
Adaptive Optimization of Wave Functions for Lattice Field Models
}

\author{
\underline{Matteo Beccaria}$^{a,b}$, Massimo Campostrini$^{a,c}$\\
Alessandra Feo$^d$
}
\instit{
$^a$Istituto Nazionale di Fisica Nucleare\\
$^b$Dipartimento di Fisica, Universit\`a di Lecce\\
$^c$Dipartimento di Fisica, Universit\`a di Pisa\\
$^d$ Institut f\"ur Theoretische Physik, \\
Westf\"alische Wilhelms-Universit\"at,
M\"unster, Germany.
}

\gdef\theauthor{M.\ Beccaria, M.\ Campostrini, A.\ Feo}

\gdef\thetitle{GFMC Calculation of Wave Functions for Quantum Field Models}

\maketitle

\begin{abstract}
The accuracy of Green Function Monte Carlo (GFMC) simulations can be
greatly improved by a clever choice of the approximate ground state
wave function that controls configuration sampling.
This trial wave function typically depends on many free parameters whose
fixing is a non trivial task. Here, we discuss a general purpose
adaptive algorithm for their non-linear optimization.

As a non trivial application we test the method on the two dimensional
Wess-Zumino model, a relativistically invariant supersymmetric field
theory with interacting bosonic and fermionic degrees of freedom.
\end{abstract}

\section{Introduction}

The traditional algorithms for numerical simulations of Lattice Field
Theories are based on the Lagrangian formulation~\cite{Lagr} that follows
Feynman's idea of representing quantum amplitudes in terms of sums
over classical paths.  An interesting alternative is the
Hamiltonian framework of Kogut-Susskind~\cite{KS} focusing on the
Hamiltonian as the generator of the temporal evolution.  
To improve numerical efficiency in the evaluation of physical
observables, analytical approximations
of the ground state wave function can be successfully exploited.
This well known procedure is called Importance
Sampling~\cite{IS} and its key problem is to determine the optimal
trial wave function (TWF) within a given class depending
parametrically on a set of free parameters.

In this contribution, 
we review a recently proposed algorithm that solves this
problem. It is built upon a standard Green Function Monte Carlo (GFMC)
simulation algorithm and is based on a non trivial feed-back between
Monte Carlo evolution and TWF modifications.

To discuss the method, we present some novel results concerning
the calculation of the ground
state energy of the supersymmetric $N=1$ Wess-Zumino model in $1+1$
dimensions.

\section{Review of Green Function Monte Carlo}

GFMC algorithms can be regarded as  numerical implementations of
the Feynman-Kac formula~\cite{FK}. In the simple case of Quantum
Mechanics, this formula claims that, 
given the potential $V(q)$ and an initial wave function
$\psi_0(q)$, the function
$$
\psi(q, t) = \ev{\psi_0(q+W_t)\exp-\int_0^t V(q+W_s) ds} ,
$$
(where $W_t$ is the Wiener process) solves the Euclidean Schr\"odinger
equation
$$
\partial_t \psi(q, t) = \frac 1 2 \Delta \psi(q, t) - V(q) \psi(q, t)
$$
with initial data $\psi(q,0) = \psi_0(q)$.

A translation of this formula into a numerical algorithm is
straightforward and quantum expectation values over the ground state
can be expressed as statistical averages over an
ensemble of weighted walkers~\cite{MB}.  As is well known, the weight
variance of the ensemble members must be controlled in some way.  Here we
adopt the Stochastic Reconfiguration algorithm with fixed population
size~\cite{Sorella}.  The finite population size bias as well as
walker correlations vanish with increasing number of walkers.

Weight fluctuations are closely related to the noise of measurements and
are due to the fact that $V$ is not constant along walker paths.
To significantly improve accuracy one needs to reduce these fluctuations.
Importance Sampling is a common strategy to achieve such a noise
reduction: the original Hamiltonian
$H = \frac 1 2 p^2+V(q)$ is transformed into
$$
\widetilde H = e^F H e^{-F} = \frac 1 2 p^2 + ip\cdot \nabla F +
\widetilde V ,
$$
$$
\widetilde V = V-\frac 1 2 \Delta F -\frac 1 2 (\nabla F)^2 .
$$
$\widetilde H$ is not canonical, but still allows simulations
with minor modifications with respect to the $F\equiv 0$
case~\cite{MB}.  The 
diffusion of the walkers is driven by a drift term proportional to
$\nabla F$ while the potential $V(q)$ is replaced by a new 
potential $\widetilde V$ that depends on the trial wave
function through $F$.  Roughly speaking, a true 
improvement is reached when $\widetilde
V$ is {\em more constant} than $V$ along
random paths in some way we are going to discuss later on.

\section{Adaptive Optimization of the Wave Function}

In this Section, we show how the trial wave function $F$ can be
optimized dynamically within Monte Carlo evolution. To this aim, we
consider a parametric TWF $\exp F(\bq, \boa)$ depending on the state
degrees of freedom $\bq$
and on a set of free parameters $\boa$.  
After $N$ Monte Carlo steps, a
simulation with a population of $K$ walkers furnishes
a biased estimator $\hat E_0(N, K, \boa)$ of the ground state energy 
$E_0$ that we shall take as a representative observable.
$\hat E_0$ is thus a
random variable such that
\be \langle \hat
E_0(N, K, \boa) \rangle = E_0 + \frac{c_1(\boa)}{K^\alpha} +
o(K^{-\alpha}), \qquad \alpha>0, \ee and \be \mbox{Var}\ \hat E_0(N,
K, \boa) \sim \frac{c_2(K, \boa)}{\sqrt{N}}, 
\ee 
where
$\langle\cdot\rangle$ is the average with respect to Monte Carlo
realizations.

In the $K\to\infty$ limit, $\langle \hat E_0\rangle$ is therefore
exact and
independent on $\boa$.  The constant $c_2(K, \boa)$ is related to the
fluctuations of the effective potential $\widetilde V$ and 
is strongly dependent on $\boa$. The problem of finding
the optimal $F$ can be translated in the minimization of $c_2(K, \boa)$
with respect to $\boa$.

The algorithm we propose~\cite{MB} 
performs this task by interlacing a Stochastic
Gradient steepest descent with the Monte Carlo evolution of the walkers
ensemble. At each Monte Carlo step, we update
$\boa_n\to\boa_{n+1}$ according to the simple law 
\be 
\boa_{n+1} =
\boa_n -\eta_n \nabla_\boa \mbox{Var}_{\ens_n}\ \widetilde V 
\ee where
${\cal E}_n$ is the ensemble at step $n$ and $\{\eta\}$ is a suitable
sequence, asymptotically vanishing.

A non-linear feedback is thus established between the TWF
 and the evolution of the walkers. 
The convergence of the method can not be easily
investigated by analytical methods and explicit numerical
simulations are required to understand the robustness of the
algorithm. 
In reference~\cite{MB}, examples of applications of the method 
with purely bosonic or fermionic degrees of freedom can be found.
In the next Section, we 
shall address a model with both kinds of fields linked together by an
exact lattice supersymmetry.

\section{The $N=1$ Wess-Zumino 
Model in $1+1$ Dimensions and Supersymmetry Breaking}

The model we consider is a lattice version of the $N=1$ Wess-Zumino
model previously studied 
in~\cite{Schiller}.  On each site of a spatial lattice with $L$ sites,
we define a real scalar field $\{\varphi_n\}$ together with its
conjugate momentum $\{p_n\}$ such that $[p_n, \varphi_m] =
-i\delta_{n,m}$. 
The associated fermion is a Majorana fermion
$\{\psi_{a, n}\}$ with $a=1, 2$ and $\{\psi_{a, n}, \psi_{b, m}\} =
\delta_{a,b}\delta_{n,m}$ , $\psi_{a,n}^\dagger = \psi_{a,n}$. The
fermionic charge
$$
Q = \sum_{n=1}^L\left [
p_n\psi_{1,n}-\left(\frac{\varphi_{n+1}-\varphi_{n-1}}{2}
+V(\varphi_n)\right)\psi_{2,n}\right],
$$
with arbitrary real {\em prepotential} $V(\varphi)$, can be used to
define a semi-positive definite Hamiltonian $H = Q^2$.  Its positive
eigenstates are automatically paired in doublets connected by $Q$;
only a $Q$-symmetric ground state with zero energy can be unpaired.
$H$ describes an interacting model, symmetric with respect to $Q$; its
ground state breaks the symmetry if and only if its energy is
positive.  We stress that spontaneous supersymmetry breaking can
occur even for finite $L$, because tunneling among degenerate vacua
connected by $Q$ is forbidden by fermion number conservation.

We can replace the two Majorana fermion operators with a single Dirac
operator satisfying $\{\chi_n, \chi_m\} = 0$, $\{\chi_n,
\chi_m^\dagger\} = \delta_{n,m}$~\cite{Schiller}.  The Hamiltonian
takes then the form \ba H &=& \sum_n\left\{ \frac 1 2 p_n^2 + \frac 1
2\left(\frac{\varphi_{n+1}-\varphi_{n-1}}{2} + V(\varphi_n)\right)^2 +
\right. \\ && \left .  -\frac 1 2 (\chi^\dagger_n\chi_{n+1} + h.c.) +
(-1)^n V'(\varphi_n) \left(\chi^\dagger_n\chi_n-\frac 1 2 \right)
\right\} \\ &=& H_B(p, \varphi) + H_F(\chi, \chi^\dagger, \varphi) \ea
and describes canonical bosonic and fermionic fields with standard
kinetic energies and a Yukawa coupling.

The theoretical problem that we want to address is dynamical breaking 
of supersymmetry in the above kind of models.
At tree level, supersymmetry breaking occurs if and only if 
$V(\varphi)$ has no zeroes.  In $1+1$ dimensions, this picture is
known to be unstable with respect to radiative corrections as
shown by the analysis of the one-loop effective potential of the scalar field
$\varphi$. 
From the non-perturbative point of view, 
the ground state energy $E_0$ is the crucial
quantity because it vanishes if and only if we are in the symmetric
phase. The choice of a Hamiltonian Monte
Carlo algorithm is then natural since $E_0$ is the simplest 
observable that can be measured with such a technique.

We study two specific examples: the cubic prepotential
$V(\varphi)=\varphi^3$, where unbroken supersymmetry is expected, and
the family of quadratic prepotentials $V(\varphi) = \lambda_0 +
\varphi^2$.  In the latter case, at tree level supersymmetry is broken
for $\lambda_0 > 0$.  Perturbative calculations and previous numerical
results~\cite{Schiller} suggest that this conclusion does not hold at
the quantum level and provide evidence that supersymmetry is broken in
the lattice model for $\lambda_0$ greater than some negative
$\lambda_0^*<0$ and is recovered for $\lambda_0\le \lambda_0^*$.  On
the other hand, the strong-coupling limit suggests $E_0>0$ for all
$\lambda_0$, decreasing exponentially for $\lambda_0\to
-\infty$~\cite{Witten}; moreover, one can show that, at fixed finite
$L$, $E_0$ is an analytical function of $\lambda_0$; this would imply
that a symmetric phase is only possible in the $L\to\infty$ limit.

\subsection{Simulation Algorithm: Brief Description}

The Monte Carlo evolution in GFMC algorithms approximates 
the Hamiltonian evolution semigroup 
$\{e^{-t H}\}_{t\ge0}$. The bosonic and fermionic 
degrees of freedom can be split by writing 
\be
\label{splitting}
e^{-\beta H} = e^{-1/2\ \beta H_B} e^{-\beta H_F} e^{-1/2\ \beta H_B}+
{\cal O}(\beta^3) , 
\ee
and the $\beta\to 0$ extrapolation is numerically computed 
at the end.
We implement the evolution associated to $H_B$ to second order in
$\beta$~\cite{Chin} and the evolution associated to $H_F$
exactly~\cite{GN}.

About the choice of the trial wave function, we propose 
the factorized form
\be
|\Psi_0^T\rangle = e^{S_B(\varphi)+S_F(\varphi, \chi,
\chi^\dagger)} | \Psi_0\rangle_B \otimes
|\Psi_0\rangle_F, 
\ee 
where $|\Psi_0\rangle_B \otimes
|\Psi_0\rangle_F$ is the ground state for the free model
and
$$
S_B = \sum_n \sum_{k= 1}^{d_B} \alpha^B_k \varphi_n^k, \quad S_F =
\sum_n (-1)^n \left(\chi_n^\dagger\chi_n-\frac 1 2\right)\
\sum_{k=1}^{d_F} \alpha^F_k \varphi_n^k.
$$
The degrees $d_B$ and $d_F$ must be chosen carefully to achieve
convergence of the adaptive determination of $\{\alpha^B, \alpha^F\}$.

In the regime of weak coupling perturbation theory,
the choice of periodic boundary conditions does not break 
supersymmetry when $L\mod 4=0$. 
Under this condition, there is an even number of fermions, $L/2$, in 
the ground state. A sign-problem then arises due to boundary crossings
since they involve an odd number of fermion exchanges.
To avoid such a difficulty, we shall adopt open boundary conditions.
With this choice, $L$ needs just to be even to assure a 
supersymmetric weak coupling ground state.
In the following, we shall restrict ourselves to the case
$L \mod 4 = 2$ and to the sector with $L/2$ fermions 
that contains a non-degenerate ground state, with zero energy
at all orders in a weak coupling expansion.

\subsection{Numerical Results}

We choose $d_B=d_F=4$;
due to the symmetries of the model with quadratic $V(\varphi) =
\lambda_0+\varphi^2$, we impose parity constraints on the trial wave
function, 
setting $\alpha^B_1$, $\alpha^B_3$, $\alpha^F_2$, $\alpha^F_4$ to zero
($\alpha^B_1$, $\alpha^B_3$, $\alpha^F_1$, $\alpha^F_3$ in the cubic
case), and and determine by the adaptive algorithm the remaining four
free parameters.
We work on a
lattice with 10 spatial sites and vary both $K$ and $\lambda_0$. In
principle, we also extrapolate any result to the $\beta\to 0$ limit.
However, the results we present are all obtained with a fixed value
$\beta=0.01$ since we checked that a reduction of this parameter by a
factor 2 does not change appreciably the results.

\begin{figure}[tb]
\centerline{\psfig{file=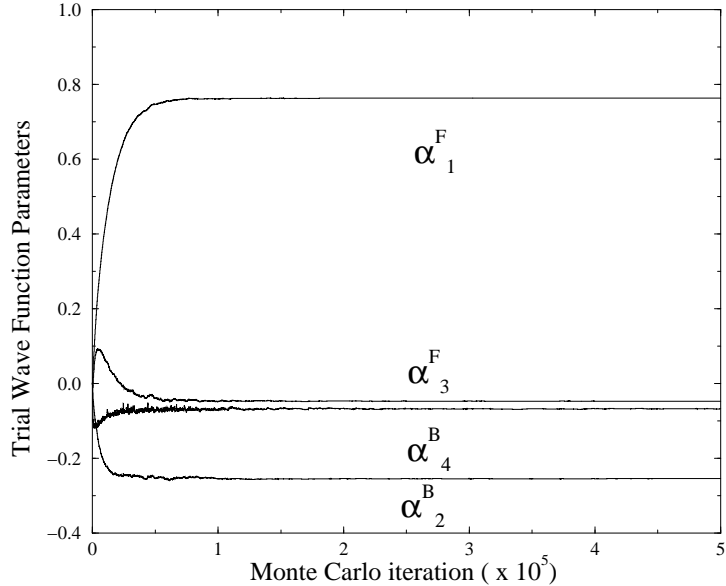,width=8cm,angle=-90}}
\caption{Adaptive Optimization of the four coefficients appearing in
the trial wave function for the quadratic model. The constant $\lambda_0$
is $0.0$ and the ensemble size is $K=100$.}
\label{Evolution}
\end{figure}

\begin{figure}[tb]
\centerline{\psfig{file=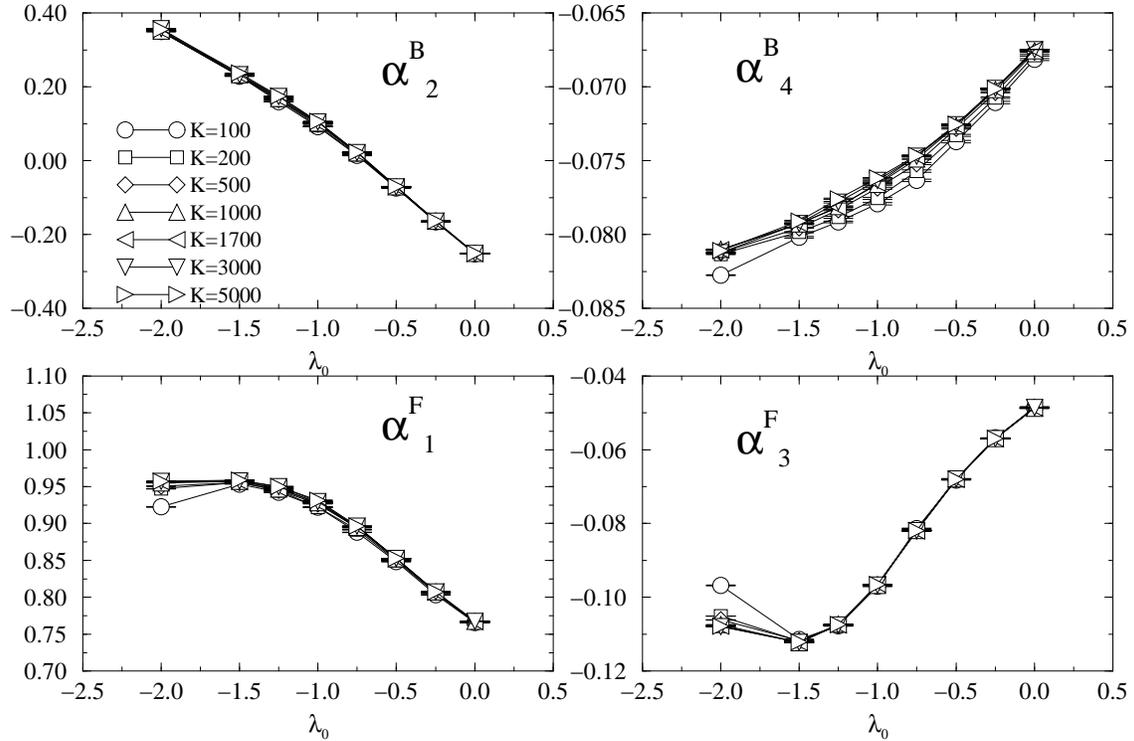,width=10cm,angle=-90}}
\caption{Summary plots of the four trial wave function parameters as
a function of $\lambda_0$ and $K$.}
\label{VariationalPars}
\end{figure}

Beginning with a small ensemble of $K=100$ walkers, we determine
adaptively the four coefficients $\{\alpha^B_2, \alpha^B_4,
\alpha^F_1, \alpha^F_3\}$ starting from zero values. A typical run is
shown in Fig.~(\ref{Evolution}).  For larger $K$, we start from the
values of $\alpha$ obtained in a run at the nearest smaller
$K$. A summary plot of $\{\alpha^B_2, \alpha^B_4, \alpha^F_1,
\alpha^F_3\}$ as functions of $K$ and $\lambda_0$ is shown in
Fig.~(\ref{VariationalPars}).

With the optimized TWF parameters, we determine
the energy contributions coming from the bosonic and fermionic sectors
separately. 
The sum of the two is the total ground
state energy $E_0$.  Apart from the dependence on $\beta$, $E_0$ is a
function of three parameters $E_0 = E_0(\lambda_0, L, K)$,
where we understand 
our choice of TWF and also optimization of its
free parameters.  The dependence on
$K$ is totally artificial and we are interested in $E_0(\lambda_0, L,
\infty)$.  The dependence on $L$ enters the study of the
continuum limit of the model.  However, here we simply want to test
the method and work at fixed $L=10$.

\begin{figure}[tb]
\centerline{\psfig{file=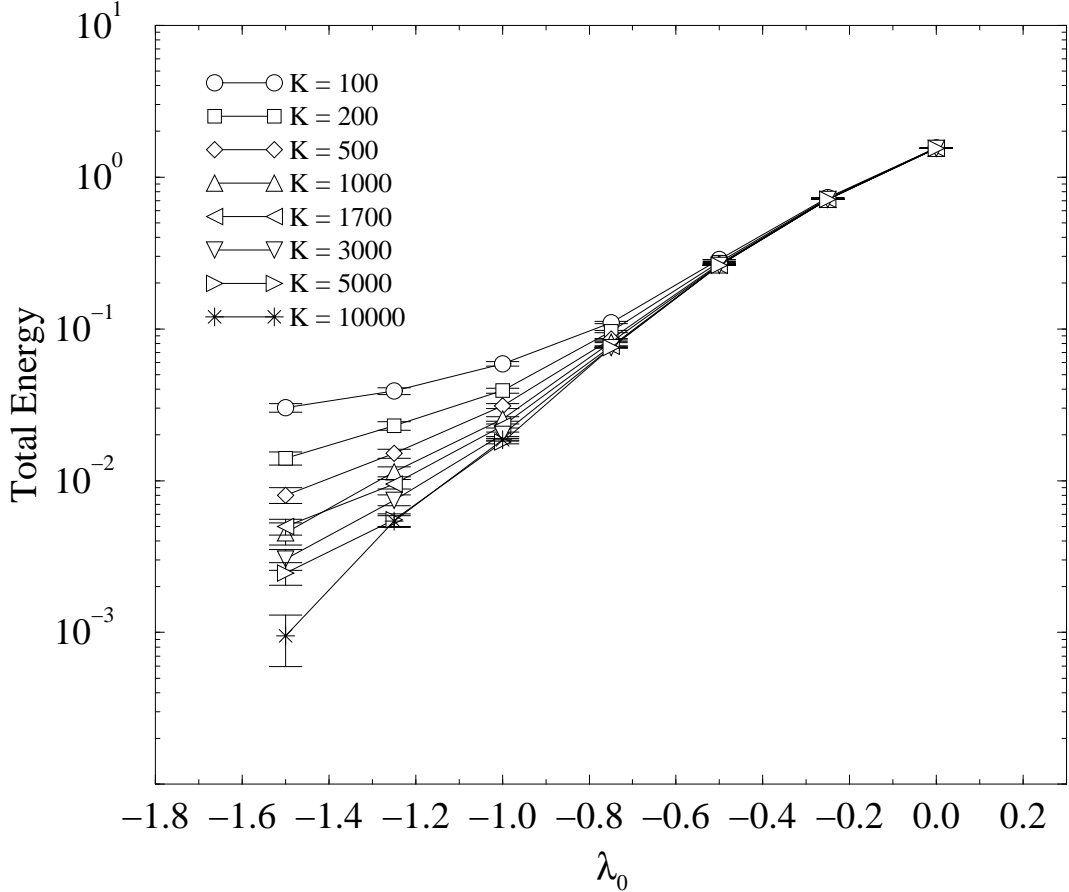,width=12cm,angle=-90}}
\caption{Total energy in log scale as a function of $\lambda_0$
and $K$.}
\label{TotalEnergy}
\end{figure}

\begin{figure}[tb]
\centerline{\psfig{file=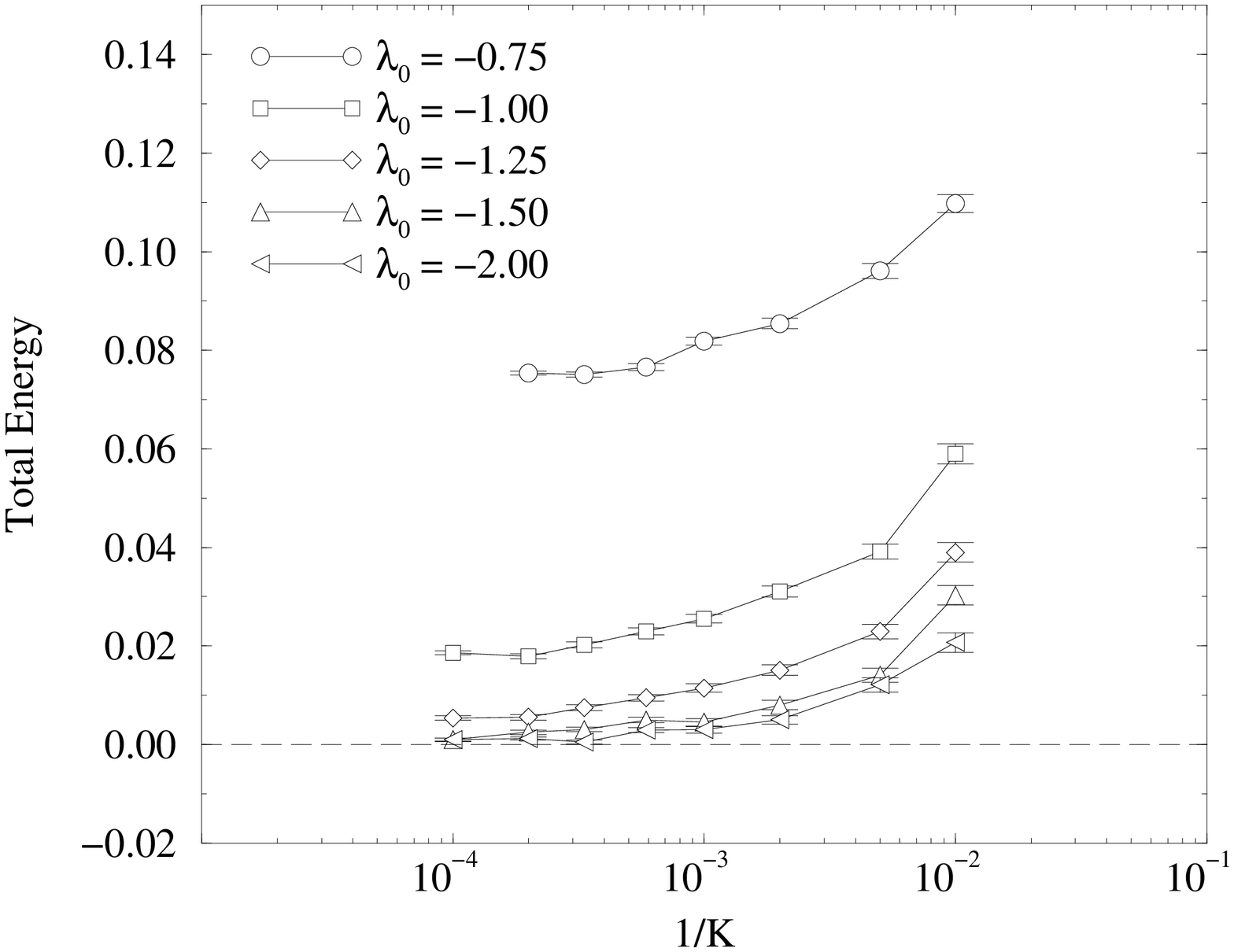,width=12cm,angle=-90}}
\caption{Extrapolation of the total energy in the $K\to\infty$
limit. The various curves are drawn for several values of
$\lambda_0$. Data points corresponds to $K=100$, $200$, $500$,
$1000$, $1700$, $3000$, $5000$, and $10000$.}
\label{TotalEnergyVersusK}
\end{figure}

For the quadratic case, we show in Fig.~(\ref{TotalEnergy}) the
behavior of $E_0(\lambda_0, L, K)$ as a function of $K$ at several
values of $\lambda_0$. For $\lambda_0>-0.5$ the $K$ dependence is very
mild and one can confirm the claim that a positive $E_0(\lambda_0, L,
\infty)>0$ is obtained at $K\to\infty$.  Instead, for $\lambda_0 <
-0.5$, there is a certain dependence on $K$ and an extrapolation
toward $K\to\infty$ must be performed to determine the asymptotic
limit.  Fig.~(\ref{TotalEnergyVersusK}) shows the results of such an
extrapolation.  For $\lambda_0 \ge -1.25$, we can exclude a zero
$E_0(\lambda_0, L, \infty)$; for $\lambda_0 \le -1.5$ our data are
compatible with zero.

\begin{figure}[tb]
\centerline{\psfig{file=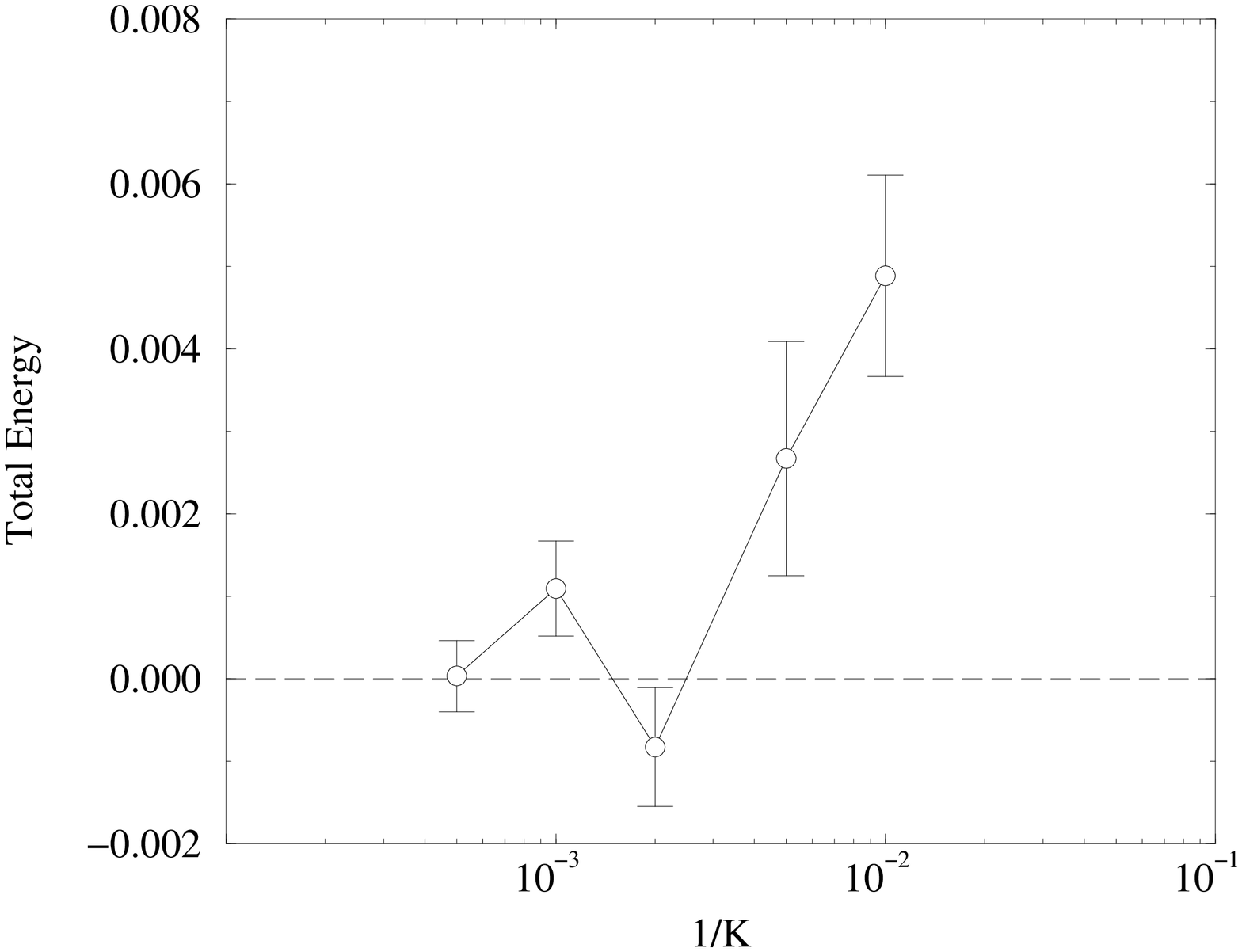,width=12cm,angle=-90}}
\caption{For the cubic potential, extrapolation of the total energy in
the $K\to\infty$ limit. Data points corresponds to $K=100$, $200$,
$500$, $1000$, and $2000$.}
\label{CubicEnergy}
\end{figure}

On the other hand, for the cubic potential the dependence of $E_0$ on
$K$ is relatively mild; in Fig.~(\ref{CubicEnergy}) we show that $E_0$
is compatible with $0$ for $K\ge500$, in full agreement with the
expectation of unbroken supersymmetry.  It should be noticed that
bosonic and fermionic contributions to $E_0$ are of the order of $10$,
and the two are canceling to a precision of $10^{-4}$.

The conclusion we can draw from the above numerical data is that
the presented algorithm behaves in a completely satisfactory way in
the analysis of the $N=1$ Wess-Zumino model, at least for what
concerns the determination of the ground state energy. In the $L=10$
lattice quadratic model supersymmetry is broken with no doubts for
values of $\lambda_0$ down to about $\lambda_0\simeq -1.25$. Below
this value, $E_0$ is compatible with zero at the statistical level we
work.  We remark that around $\lambda_0\simeq -0.75$ the coefficient
$\alpha_2^B$ changes sign, modifying the shape of the TWF and allowing
for local minima in its $\varphi$-dependent part $S_B(\varphi)$ at non
zero fields; we interpret this fact as an interesting signal that some
transition is occurring and emphasize the important role that trial
wave functions play.  On the other hand, for the cubic model the
evidence for unbroken supersymmetry is quite convincing.

Unbroken supersymmetry implies a number of non-trivial Ward
identities; we monitored several of these, obtaining a pattern of
supersymmetry breaking perfectly consistent with the one obtained from
$E_0$, although with lower numerical precision~\cite{Lattice}.

\section{Conclusions}
 
The Hamiltonian approach is a powerful method for Monte Carlo analysis
of field models. It is well founded and general purpose, but the 
control of systematic errors is fundamental. A good trial wave function can
improve strongly the quality of numerical results and the
convergence rate of simulations.  The trial wave function is an
approximation to the exact ground state. As such, it contains
important physical informations about the model under study.
Optimization of (many-parameter) wave functions is thus not only
motivated by computational needs only. It affords and checks
analytical insights about the actual ground state.

Moreover, as we have shown, for certain models in $1+1$ dimensions,
the Hamiltonian formalism appears to be the natural framework for
lattice fermions.  The treatment of bosons and fermions is nicely
symmetric and the non-local determinants required in Lagrangian
simulations are not required at all.  The possibility of preserving
exactly a 1-dimensional supersymmetry algebra is also very appealing.


\end{document}